# Audience Reach of Scientific Data Visualizations in Planetarium-Screened Films

Kalina Borkiewicz, University of Utah and University of Illinois at Urbana-Champaign
Eric Jensen, University of Illinois at Urbana-Champaign
Yiwen Miao, University of Illinois at Urbana-Champaign
Stuart Levy, University of Illinois at Urbana-Champaign
J.P. Naiman, University of Illinois at Urbana-Champaign
Jeff Carpenter, University of Illinois at Urbana-Champaign
Katherine E. Isaacs, University of Utah

**Abstract**

Quantifying the global reach of planetarium dome shows presents significant challenges due to the lack of standardized viewership tracking mechanisms across diverse planetarium venues. We present an analysis of the global impact of dome shows, presenting data regarding four documentary films from a single visualization lab. Specifically, we designed and administered a viewership survey of four long-running shows that contained cinematic scientific visualizations. Reported survey data shows that between 1.2 - 2.6 million people have viewed these four films across the 68 responding planetariums (mean: 1.9 million). When we include estimates and extrapolate for the 315 planetariums that licensed these shows, we arrive at an estimate of 16.5 - 24.1 million people having seen these films (mean: 20.3 million).

## 1. Introduction

Immersive experiences in planetariums, museums, science centers, and other educational environments with domed theaters (collectively referred to as "planetariums" hereafter) strive to engage diverse audiences through a mix of live presentations and pre-rendered dome show films. These films often incorporate cinematic scientific visualizations (Borkiewicz 2020), offering an impactful way to communicate complex scientific findings to the public. This study focuses on pre-rendered dome shows, particularly those that leverage the power of scientific visualization to enhance educational content. The growing importance of cinematic scientific visualization in planetarium settings was underscored in the 2023 IEEE VIS keynote, "Visualizing The Chemistry of Life on Giant 360-degree Screens" (Berry and Ynnerman 2023). Dome shows, and the scientific visualizations they contain, can have a profound global reach. As our data reveals, many dome shows remain in circulation for decades and continue to be screened long after their initial release. By analyzing the reach and viewership of four specific dome shows from a single visualization lab, this paper aims to shed light on their global impact.

Considering the substantial investments of time, effort, and funding required to create cinematic visualizations, it is imperative to quantify their impact. Understanding the reach of these visualizations is essential for assessing their potential value and determining whether the investment in their creation is justified. However, there are several challenges to determining the precise number of viewers for any given dome show. Unlike traditional film screenings where ticket sales provide concrete metrics, planetariums lack standardized mechanisms for tracking viewership. Moreover, planetariums range from small portable facilities to major institutions up to 30 meters in diameter (Yu et al 2016). While major planetariums may maintain detailed attendance records, many smaller venues lack such infrastructure, posing significant challenges in obtaining comprehensive viewership data. Small domes are much more prevalent than large domes globally (Petersen 2020) and thus contribute a sizable portion of all dome show audiences, making it imperative that these venues are included in analyzing viewership data.

To evaluate the global reach of planetarium dome shows containing visualizations, we surveyed the planetariums that licensed four such films. Based on the survey responses, we determine a minimum viewership of 1.2 million. We then extrapolate from the list of total licensees to provide a low-range estimate of 16.5 million viewers. Our survey and analysis processes, along with detailed results, is presented below.

## 2. Data Collection and Analysis

Four dome documentary films were selected for this study: *Birth of Planet Earth* (2019), *Solar Superstorms* (2015), *Dynamic Earth* (2012), and *Black Holes: The Other Side of Infinity* (2007). These films feature visualizations from the Advanced Visualization Lab at the National Center for Supercomputing Applications at the University of Illinois at Urbana-Champaign. We chose these films because they contain cinematic scientific visualizations and licensing data was available to the authors. We sent a survey regarding viewership and received responses from 23% of the total licensees. We then deductively coded the data for viewership as described in Section 2.1. Finally, we combined the data with information about planetarium sizes to estimate viewership across all licensees.

### 2.1. Data Compilation

Evans and Sutherland[1], the films' distributor, provided the sales data for the four selected films. The data included the name of each planetarium, city, length of the purchased license, and purchase date. We coalesced data by planetarium, i.e. if a single planetarium re-purchased a show after the expiration of a license, we combined rows, using the combined length of the purchased licenses and first purchase date. The resulting data contains information for 315 planetariums.

---

[1] https://shows.es.com/

We matched the sales data against two planetarium databases, the World Planetariums Database (Audeon and Ruiz n.d.) and the Loch Ness Productions Fulldome Theater Compendium (Petersen and Petersen n.d.), to obtain contact information for the planetariums as well as information about their sizes and capacities. The combined dataset includes the following data: (1) the seating capacity, (2) the size of each planetarium dome in meters, and (3) the annual attendance for each venue. In instances where this information was not available, we conducted a manual search to find the information on the planetariums' websites, annual reports, and/or Wikipedia pages, as applicable. We acquired dome sizing information for 280 of the planetariums which had purchased the film.

Each planetarium was categorized into one of seven size buckets, as defined by Petersen (Petersen 2020), which we label as XXS, XS, S, M, L, XL, and XXL for readability, with "U" for planetariums of unknown size. We used the dome size in meters to make this classification when it was available, otherwise, we estimated the classification using the dome capacity. When the planetarium had multiple domes, we used the largest dome size for the classification.

**2.2 Survey Data Collection**

We sent our survey to the 268 planetariums for which we could find contact information and received 72 responses. Figure 1 shows the distribution of planetariums by dome size and by film. For each purchased film, the survey participants were asked to provide data or estimates regarding the number of people who attended film screenings and information about how the data was gathered or estimated. If the planetariums did not have the requested information on hand, additional questions about their planetarium and their events were asked to assist in making estimates.

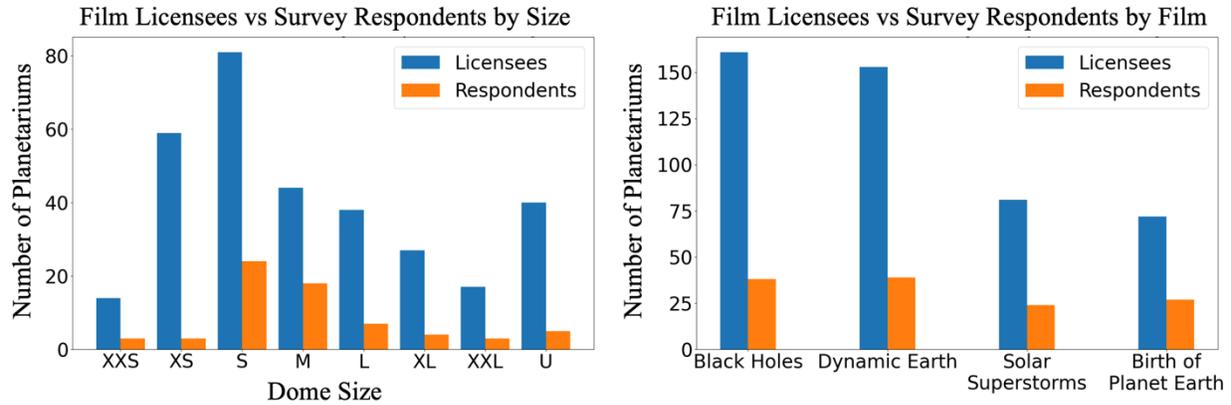

*Figure 1. Distribution of planetariums that have licensed the four studied films, compared to those who answered our survey.*

Several planetariums answered survey questions about films which they should not have had according to the licensing data from the distributor. We cross-checked their access to these films by checking listings on the planetariums' websites as well as sending follow-up emails, and we verified all but two survey answers. All of these responses were kept, and marked as licensed. Ten planetariums did not list their institution name in the survey. Where possible, we mapped these responses to planetariums by using identifying information from their responses. For example, a survey respondent that specified a seating capacity of 77, a specific and unique value, was easily matched to a corresponding planetarium among the list of purchasers. Likely matches were found for five of the ten unspecified planetariums, and the others were marked "U". One planetarium had two different people respond with slightly different answers, so we averaged them. We removed four responses for having no estimate (2) or no record (2) of screenings. After this initial data cleaning, 68 survey responses from planetariums remained. Several respondents provided information about multiple films, resulting in 105 data points for dome show viewership out of 459 license purchases across the four films.

All survey questions were optional and allowed numerical or free-text answers to encourage the submission of partial information if available. Many planetariums provided rough, text-based answers (e.g. "about 500 per year since the purchase date"), which were coded by two of the authors into a range of possible values. The coders followed a mutually agreed upon set of rules for coding, and met to discuss differences in interpretations. Where a consensus could not be reached among the coders, the lowest-low and highest-high were used for the range. We applied the following coding criteria with equations for hedge word estimation based on work from Ferson et al (Ferson et al 2015).

1. If an exact number is provided, use that value.
2. If an estimated number is provided, use the range $x \pm 2*10^{-d}$ where d is the decimal place of the last significant digit of x.
3. If a phrase like "about x" is used, use the range $x \pm 2*10^{-d}$.
4. If a phrase like "above x" is used, use the range $[x, x+ 2*10^{-d}]$.
5. If the above calculations result in numbers that are negative or zero, reduce d by one.
6. If viewership information is not provided, calculate low and high estimates using numShowsPerWeek * numWeeksPerYear * numYearsSinceLicense * seatingCapacity if these values are available.
    a. numShowsPerWeek uses the reported survey value for the "high" estimate calculation and 1 for the "low" estimate calculation. We chose this range because upon cross-checking against their websites, it appeared several planetariums misunderstood the survey question "What is the number of showings for this show per week?" to mean all shows rather than the particular title.
    b. numWeeksPerYear uses 49 ("low") and 52 ("high") weeks per year to account for planetarium closings. If the planetarium is part of a school district, the number of weeks in the school year is used.
7. Use the planetarium attendance and capacity information provided in the survey over information pulled from external databases. Use the numbers provided by respondents where possible.
8. If it is not possible to determine a number based on the response, mark the data for non-inclusion in the analysis.

Each coder coded the data independently. There were 39 differences in interpretations prompting discussion. The most common issues (27 out of 39) were easy to resolve due to a simple misapplication of a rule or a coder not utilizing all relevant information, however several instances were difficult to resolve and required turning to outside sources of information.

One large planetarium claimed to have "63+" showings of *Black Holes* per week, adding that "Attendance is strong when BH is on the schedule". This was an instance where the "63+" survey answer actually refers to the total number of shows played per week, and not specifically *Black Holes* shows. Our attempts to contact the planetarium to acquire scheduling information were unsuccessful, so per coding rule 6a, we initially coded a viewership range assuming the film is shown between 1 and $63 + 2*10^0$ times per week, which resulted in a viewership estimate of 51,000 - 3,529,000 for this planetarium---an extraordinarily unhelpful range. Coder #2 identified historical versions of the planetarium's website between 2012 - 2020 using The

Internet Archive Wayback Machine[2], and from this was able to determine that the film was actually shown between 4 times a week 20% of the time, and 3 times a day 35% of the time.

A spreadsheet detailing the licensing information, dome size information, and coding results is available at doi.org/10.7278/S5d-1anf-0dcr

**2.3. Survey Results**

The total reported number of viewers of these four films across 68 planetariums (out of 315 total) lies between 1.2 - 2.6 million (mean: 1.9 million) based on the coded survey responses: 742K - 1.81M (mean: 1.28M) for *Black Holes: The Other Side of Infinity* (2007), 184K - 406K (mean: 295K) for *Dynamic Earth* (2012), 43K - 80K (mean: 62K) for *Solar Superstorms* (2015), and 190K - 274K (mean: 232K) for *Birth of Planet Earth* (2019). The breakdown per film and per dome size classification is shown in Figure 2. This holistic analysis demonstrates that the four films together have reached millions of viewers.

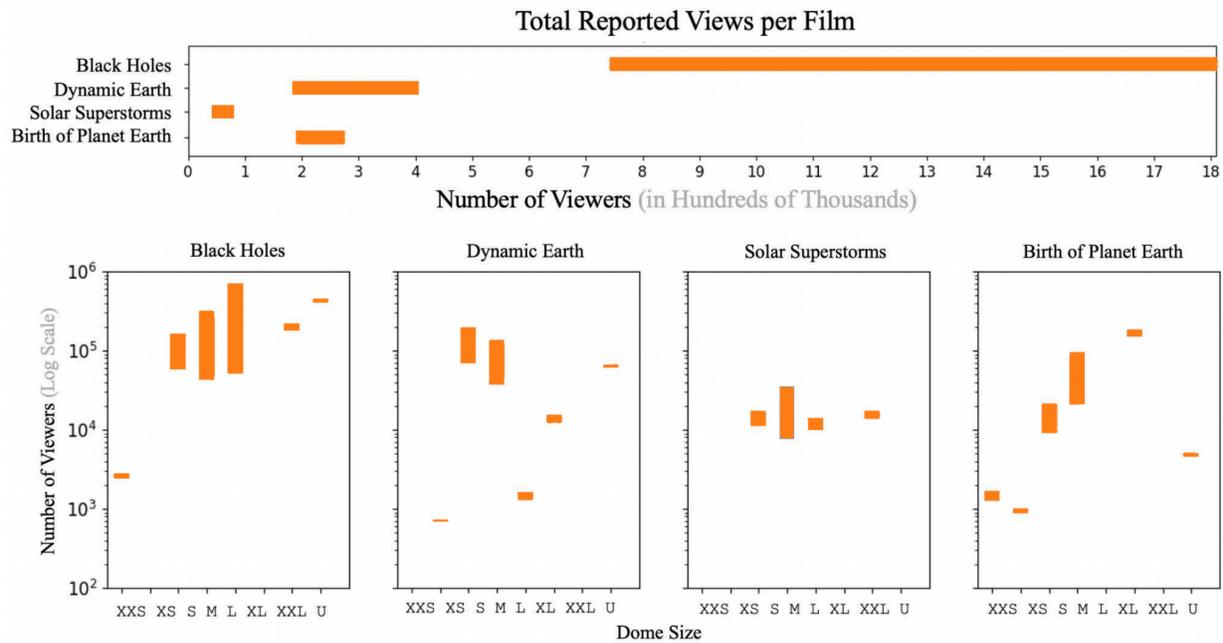

*Figure 2. Results from the 68 planetariums that answered the survey. These charts show total reported views per film, presented in log scale on bottom.*

---

[2] https://archive.org/

We report a wide range for the viewership of *Black Holes: The Other Side of Infinity* due to significant uncertainty in a small number of the coded survey responses. A z-score analysis of the variation between the lower and upper bounds identified two outliers with value ranges that fall more than 3 standard deviations from the mean. Removing these two outlier survey responses narrows the range for Black Holes to 672K - 823K (mean: 748K). As a result, the total viewership estimate for the four films also decreases to 1.1 million - 1.4 million (mean: 1.2 million).

We next consider the annual viewership of each film, considering factors like years since release and planetarium size. We will use this metric to help estimate viewership across non-responding planetariums. Specifically, we consider two interpretations of the term 'annual': (A) the average number of viewers per year since the film's release, and (B) the average number of viewers per year under a planetarium's licensing agreement. For example, in the second interpretation, a film that is 10 years old might have only been licensed for 5 of those years, affecting its annual viewership average.

To address interpretation A, each coded survey response is divided by the number of years since the films' release date. The release month and year were obtained from the Fulldome Database[3], using the 15th of the month as the release day. From this perspective, the total reported viewers annually of these four films across 68 planetariums lies between 138K - 264K (mean: 201K) since their release dates based on survey responses: 49K - 120K (mean: 84K) for *Black Holes*, 18K - 40K (mean: 29K) for *Dynamic Earth*, 6K - 12K (mean: 9K) for *Solar Superstorms*, and 64K - 93K (mean: 78K) for *Birth of Planet Earth*. The per-film and per-dome size breakdown of results is shown in Figure 3 in green. Removing the two *Black Holes* outliers reduces the film's range to 44K - 54K (mean: 49K) and the overall range to 129K - 176K (mean: 152K).

---

[3] https://www.fddb.org/

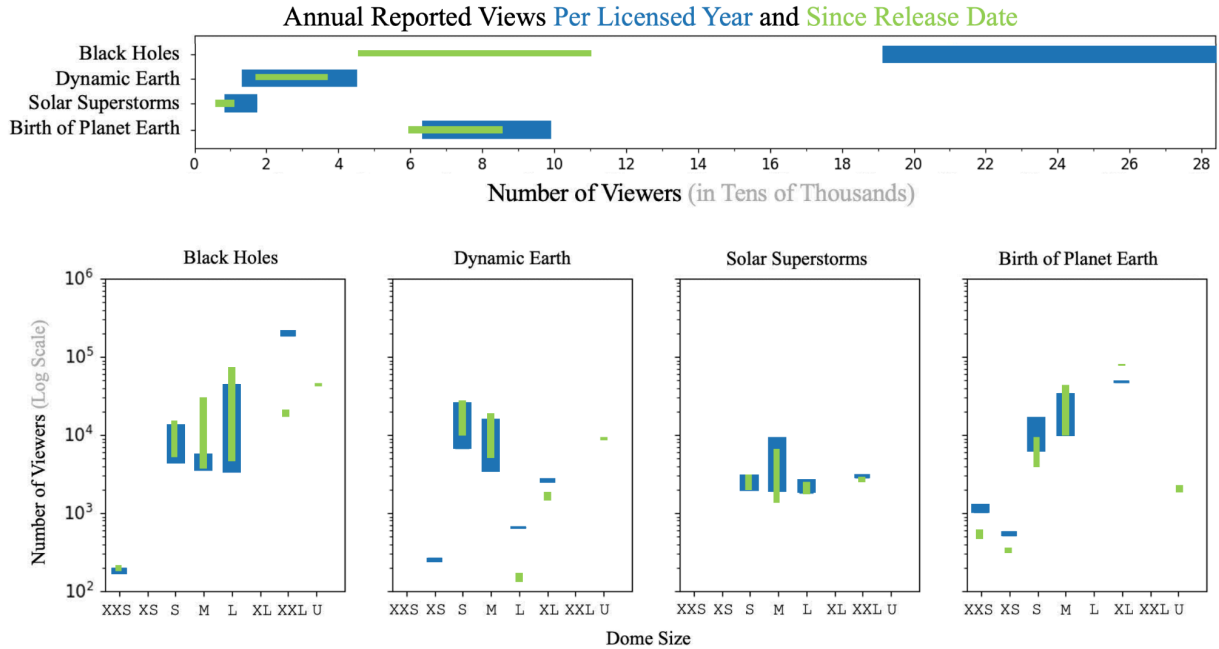

*Figure 3. Results from the survey, showing the annual number of views. The blue bars represent views in the years that the films have been licensed, the green bars represent views since the films' release.*

Interpretation B requires dividing each coded survey response by the number of years that the particular planetarium has had the film. We were unable to make this calculation for ten survey respondents due incomplete data regarding licensing years. A few of our respondents either (1) did not provide their name in the survey and could not be mapped to licensing information, or (2) responded regarding a license missing in the licensing database. Therefore, of the 68 survey responses and 105 data points from the survey, only 58 planetariums and 79 data points were usable for this calculation.

From a licensing perspective (interpretation B) for planetariums that have had a show for at least 6 months, the total annual reported viewers of these four films lies between 277K - 445K (mean: 361K) per license year across the 58 planetariums. This breaks down as 191K - 284K (mean: 238K) for *Black Holes*, 13K - 45K (mean: 29K) for *Dynamic Earth*, 8K - 17K (mean: 13K) for *Solar Superstorms*, and 64K - 99K (mean: 81K) for *Birth of Planet Earth*. The per-film and per-dome size results breakdown is shown in Figure 3 in blue. Removing the two *Black Holes* outliers reduces the film's range to 188K - 239K (mean: 214K) and the overall range to 273K - 401K (mean: 337K).

## 3. Estimating Across All Licensees

Section 2 presented viewership from the planetariums that answered our survey. We now use the data to estimate viewership across all 315 planetariums that licensed the four dome shows.

## 3.1. Data Extrapolation Method

For planetariums that did not respond to our survey, we estimated their viewership using the yearly ranges of responding planetariums in their size class for each film. Specifically, for each of the planetariums that licensed a show, and for each show they licensed, a low estimate and a high estimate was extrapolated based on the annual viewership per licensed year for their dome size. For survey responders, we used their estimates as per Section 2.

Thus, data from 58 planetariums about viewership at 79 instances of licensed dome shows was extrapolated to 315 planetariums with 459 instances of licensed dome shows.

When we did not have survey information for a particular combination of film and dome size, we used averages for neighboring size classes when possible. For example, there was no survey data for *Birth of Planet Earth* from planetariums of size "L", so the data for *Birth of Planet Earth* from planetarium sizes of "M" and "XL" were averaged.

For missing XXS and XXL data, rather than averaging based on similar sizes for that film, we averaged the data from that exact size class, across all films. For instance, there was no survey data for XXS planetariums for the film *Dynamic Earth*, so the survey data for XXS planetariums which had the three other films was averaged, as differences between the films striated by dome size were much smaller than across all sizes. When the planetarium size was unknown (U), we used the median from all planetarium sizes which had that film. Similarly, if a film was licensed for an unknown number of years, we used the median overall license length for that film.

## 3.2. Extrapolation Results

We estimate that the total reach of these four films is between 16.5 - 24.1 million (mean: 20.3 million) viewers: 12.8 - 19.4 million (mean: 16.1 million) for *Black Holes* (2007), 1.2 - 1.8 million (mean: 1.5 million) for *Dynamic Earth* (2012), 182K - 303K (mean: 243K) for *Solar Superstorms* (2015), and 2.4 - 2.6 million (mean: 2.5 million) for *Birth of Planet Earth* (2019), as shown in Figure 4. Removing the two *Black Holes* outliers reduces the film's range to 1.2 million

- 1.5 million (mean: 1.4 million) and the overall range to 15.9 million - 20.0 million (mean: 18.0 million).

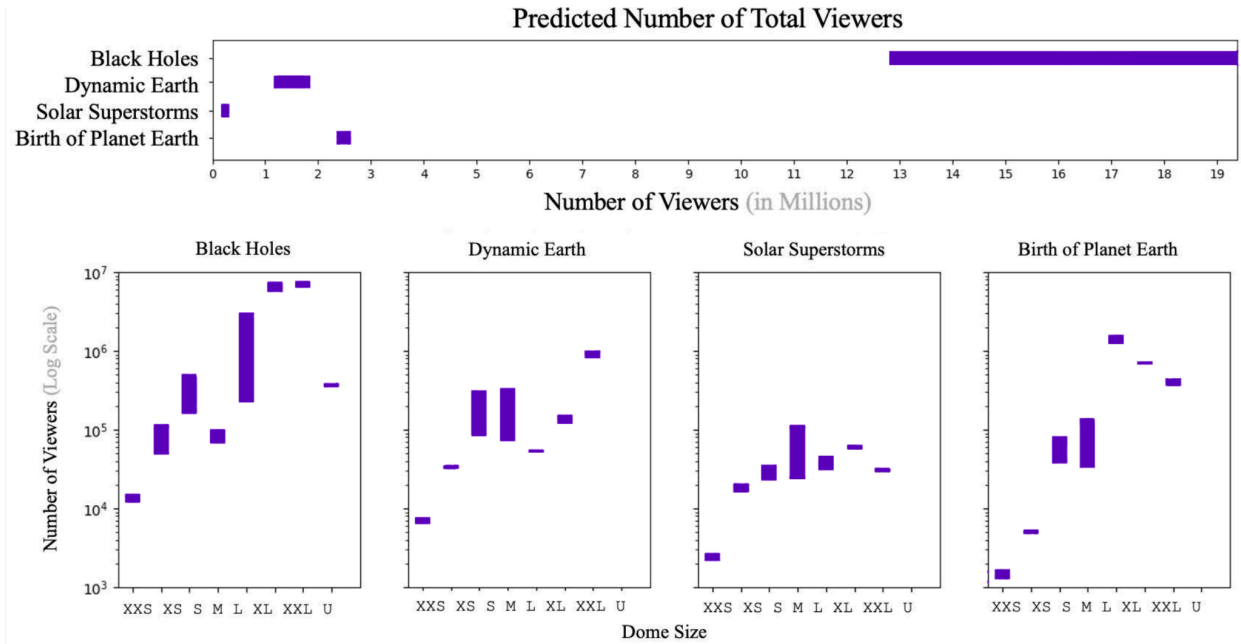

*Figure 4. Predicted number of total viewers based on extrapolation.*

Taking into account the number of years that each film has been released, the annual viewership since release date of the four films is 1.8 - 2.4 million (mean: 2.8 million): 847K - 1.28M (mean: 1.1 million)for *Black Holes*, 116K - 182K (mean: 149K) for *Dynamic Earth*, 27K - 45K (mean: 36K) for *Solar Superstorms*, and 797K - 883K (mean: 840K) for *Birth of Planet Earth*, shown in Figure 5. Removing the two *Black Holes* outliers reduces the film's range to 1.2 million - 1.5 million (mean: 1.4 million) and the overall range to 15.9 million - 20.0 million (mean: 18.0 million).

If we consider the total number of years a film will be screened, we can estimate a lifetime audience for a film. We use the term "lifespan" for the number of years a film is licensed past its release year. Jensen et al. (Jensen et al 2022) suggested that the total lifespan of these dome show films was over ten years. We use the license data provided by Evans and Sutherland to refine this estimate.

Figure 5 shows the license horizon for the four films, with *Black Holes* and *Dynamic Earth* licensed for over 50 years past their respective releases. Considering that these four films are still being actively licensed, it is likely these values are underestimated. Purchases of 10-25 year licenses of *Black Holes* still occur even 15 years after the film's release.

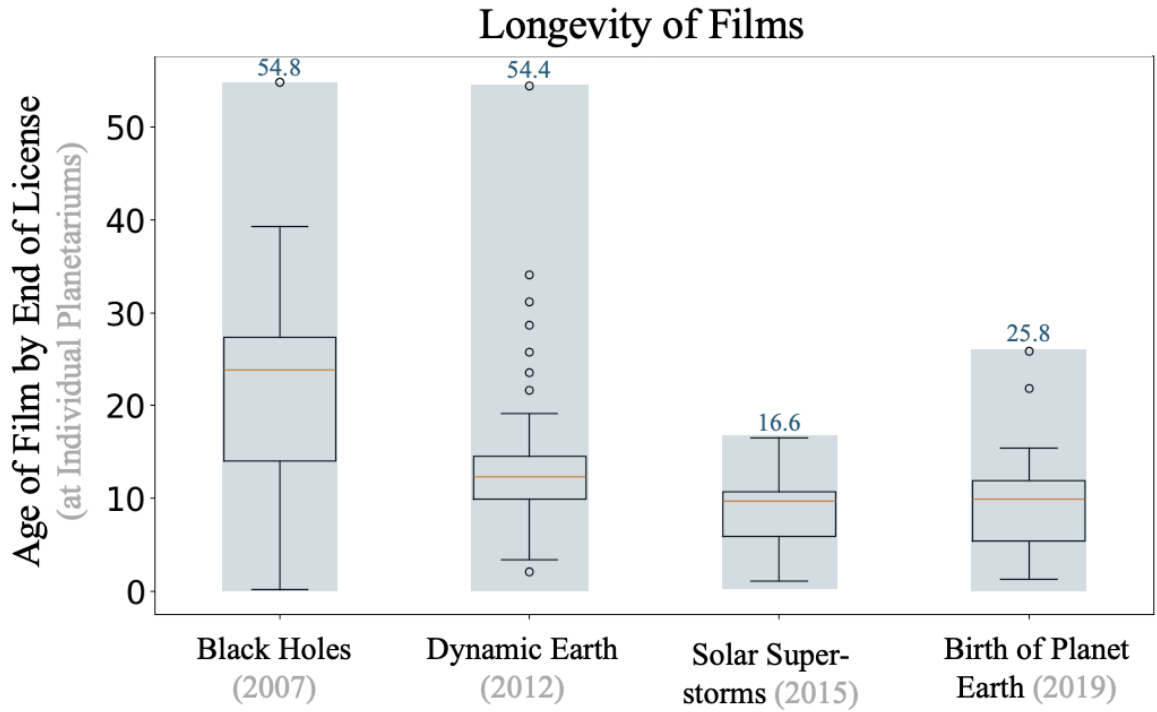

*Figure 5. Our data shows that the longevity of dome shows is much longer than the previously-claimed 10+ years (Jensen et al 2022). On the low end of these box plots, a planetarium might purchase a film on its release date and play it for one year. On the higher end, a planetarium may acquire a film 10 years after its release, and purchase a 25-year license. Two outliers show that by the time current licenses expire, some planetariums retain rights to screen 50+ year-old films.*

As we cannot predict if planetariums will use their licenses 50 years in the future, we use the conservative estimate of 10 years from Jensen et al. and average across the four films with respect to the low and high estimates for annual viewers since released. This calculation results in an expected reach of 4.9 - 6.5 million viewers (mean: 5.7 million) of a planetarium dome show over a 10-year lifespan, and the cinematic scientific visualizations therein.

**4. Limitations**

This work has several limitations. The survey was conducted in English and this is likely the reason for poor representation from countries where the primary language is not English. Of the 68 usable survey responses, 51 were from the USA, 6 from Europe, 2 from Canada, 2 from the Middle East, 1 from Australia, 1 from South America, and 5 from unknown locations. This under-representation of international planetariums likely biases the results.

Furthermore, a possible issue with *Black Holes* is that the topic of black holes is popular among dome shows, so there is a possibility that a planetarium might have confused this film with another of a similar title. To minimize this risk, when a planetarium claimed to have the film but was not represented in the licensing database, we followed up with respondents to confirm the data corresponded to "the 2007 dome film *Black Holes: The Other Side of Infinity*, narrated by Liam Neeson'. All confirmations were positive.

We also note that not all films are created equal and confounding factors such as popularity of the topic (Zajan and Stengler 2021) and quality of the film also affect viewership. For instance, black holes became a topic of wide interest when the first photograph of a black hole made international news in 2019. This press interest may have disproportionately affected the viewership for the film on that topic. Viewership results may be different from a repeated survey focusing on a different selection of films.

Moreover, many planetariums were closed during the COVID pandemic, which likely led to lower attendance in recent years. Recent viewership totals may be smaller than they would have been under normal circumstances.

Visualizations created for planetarium dome shows are often re-used in a flat-screen format and shared with the public on numerous platforms. For instance, versions of *Solar Superstorms* and *Birth of Planet Earth* are available on Amazon Prime[4,5], and a single YouTube video[6], which repurposes visualizations from *Solar Superstorms* has 3.4 million views as of this writing. Thus, the reach of a cinematic scientific visualization may be an order of magnitude greater than the estimates provided in this paper. Future work could study the reach of cinematic scientific visualizations more broadly, beyond the context of only planetariums.

Throughout our analysis, we aimed to underestimate to avoid over-reporting. For example, we found from our survey and subsequent internet searches that some planetariums were screening films not in the license database, but only used the license information to estimate viewership from non-respondents. Thus, there may be more screenings we did not account for due to incomplete licensing information.

---

[4] https://www.amazon.com/Birth-Planet-Earth-Thomas-Lucas/dp/B07W89HPNQ
[5] https://www.amazon.com/Solar-Superstorms-Journey-Center-Sun/dp/B015R1BI0M
[6] https://www.youtube.com/watch?v=inuCAqj8UgQ

## 5. Discussion and Future Work

Our survey highlights the enormous potential that cinematic scientific visualizations have for engaging the public in planetariums on a global scale. Just four visualization projects from one visualization lab have reached millions of viewers. At the most conservative estimate, only including the responses with specific estimates (68 out of 315 planetariums which provided survey responses about at least one of the four films), we can claim that 1.2 - 2.6 million people have seen the films (mean: 1.9 million). Extrapolating to all 315 planetariums and their 459 licenses in our database, we estimate a global reach of about 16.5 - 24.1 million (mean: 20.3 million). If these four films can be considered representative of all dome shows, then a producer can expect a single dome show to reach 4.9 - 6.5 million people over a 10-year lifespan (mean: 5.7 million). These are conservative estimates as films may be licensed for over 50 years.

Based on annual viewership, *Black Holes: The Other Side of Infinity* (2007) and *Birth of Planet Earth* (2019) are the most popular films from this study, and *Solar Superstorms* (2015) is the least popular, by a wide margin. Surprisingly, the star-power of *Solar Superstorms*'s narrator, Benedict Cumberbatch, did not make the film a bigger draw than *Birth of Planet Earth* which is narrated by Richard Dormer, a comparatively less-known actor. Considering that a celebrity narrator can have a large impact on the overall budget of a film, future work could study whether celebrity narration contributes to viewership, enjoyment, trust, or other measurable metrics for dome show films.

We studied planetarium shows that combine cinematic scientific data visualization with animated illustrations, recorded videos, and other media. Further work is needed to assess the impact of the data visualizations with respect to the other parts of the films on factors like audience comprehension, memorability, and trust. Investigating audience response across these mediums along with factors that amplify or attenuate their differences would aid in effectively creating future such films.

**Acknowledgements**

The authors wish to thank Jenn Davis, Carolyn Petersen, Ayesha Kesharia, Zoey Park, and Sebastian Frith.